\documentclass[english,aps,manuscript]{revtex4}
\usepackage[T1]{fontenc}
\usepackage[latin9]{inputenc}
\usepackage[a4paper]{geometry}
\usepackage{mathrsfs}
\usepackage{amsmath}
\usepackage{amssymb}
\usepackage{graphicx}
\usepackage{esint}

\makeatletter
\@ifundefined{textcolor}{}
{%
 \definecolor{BLACK}{gray}{0}
 \definecolor{WHITE}{gray}{1}
 \definecolor{RED}{rgb}{1,0,0}
 \definecolor{GREEN}{rgb}{0,1,0}
 \definecolor{BLUE}{rgb}{0,0,1}
 \definecolor{CYAN}{cmyk}{1,0,0,0}
 \definecolor{MAGENTA}{cmyk}{0,1,0,0}
 \definecolor{YELLOW}{cmyk}{0,0,1,0}
}

\usepackage{mathrsfs}

\@ifundefined{textcolor}{}{%
 \definecolor{BLACK}{gray}{0}
 \definecolor{WHITE}{gray}{1}
 \definecolor{RED}{rgb}{1,0,0}
 \definecolor{GREEN}{rgb}{0,1,0}
 \definecolor{BLUE}{rgb}{0,0,1}
 \definecolor{CYAN}{cmyk}{1,0,0,0}
 \definecolor{MAGENTA}{cmyk}{0,1,0,0}
 \definecolor{YELLOW}{cmyk}{0,0,1,0}
}

\usepackage{babel}

\usepackage{babel}

\makeatother

\usepackage{babel}
\begin{document}

\title{Consensus formation times in fully connected societies }

\author{Juan Neirotti}

\address{Department of Mathematics, Aston University, The Aston Triangle,
B4 7ET, Birmingham, UK}
\begin{abstract}
We developed a statistical mechanics approach to the problem of opinion
formation in interacting agents, constrained by a set of social rules,
\emph{B}. To provide the agents with an adaptive quality, we represented
both the social agents and the social rule by perceptrons. For fully
connected societies we find that if the agents' interaction is weak,
all agents adapt to the social rule \emph{B}, with which they form
a consensus; but if the interaction is sufficiently strong a consensus
is built against the established \emph{status quo}. This behavior
is observed for all temperatures $T$ and for all values of the agents'
interaction parameter $H_{0}$, except in the limit $T\to\infty$
or when the interaction reaches the critical value $H_{0}=1,$ where
no consensus is formed. The agents follow a path where, after a time
$\alpha_{c},$ they disregard their peers' opinions on socially neutral
issues and reach a full consensus at time $\alpha_{d}>\alpha_{c}.$
The measure of time $\alpha$ is proportional to the volume of information
provided to the agents. 
\end{abstract}
\maketitle
In this letter we propose a statistical mechanics approach to study
the emergence of consensus in a fully connected society of adaptive
agents, in the presence of a social field $B.$ The term \emph{consensus}
is understood as the level of agreement amongst the agents in favor
or against the predetermined socially accepted position delivered
by $B$ \cite{torok}. $B$ represents the set of rules resulting
from previous consensus-forming processes, typically observed in any
functioning society \cite{galam1,galam2}. Agents form their opinions
on social issues based on partial information received regularly during
the process. The volume of information increases over time and, the
agents being adaptive, they update their opinions accordingly. At
the end of the process the rate of agreement between agents and $B$
is measured to determine whether a consensus is formed supporting
or rebutting the social order. 

There is sufficient evidence in support of modeling opinions (on \emph{important}
issues) with binary variables \cite{kacperski}. We will represent
the opinion of agent $a$ on an issue $\boldsymbol{\xi}\in\{\pm1\}^{N}$
(represented by a binary string of length $N$) by $\sigma_{a}(\boldsymbol{\xi})\in\{\pm1\}$.
In mathematical terms $B$ is a classifier that assigns a binary label
$\sigma_{B}(\boldsymbol{\xi})$ to the issue $\boldsymbol{\xi}$.
According to \cite{neirotti1}, representing $B$ with a perceptron
(with a constant synaptic vector \textbf{$\mathbf{B}\in\mathbb{R}^{N}$})
ensures the analytical tractability of the model. In this manner the
socially accepted position on $\boldsymbol{\xi}$ is $\sigma_{B}(\boldsymbol{\xi})=\mathrm{sgn}({\bf B}\cdot\boldsymbol{\xi})$
where $\mathrm{sgn}(x)=1$ if $x>0,$ $-1$ if $x<0$ and 0 otherwise
and ${\bf B}\cdot\boldsymbol{\xi}=\sum_{j=1}^{N}B_{j}\xi_{j}$. For
consistency sake we associate each agent $a$ to a perceptron with
an adaptive synaptic vector ${\bf J}_{a}$, such that $\sigma_{a}(\boldsymbol{\xi})=\mathrm{sgn}({\bf J}_{a}\cdot\boldsymbol{\xi})$.

There is a body of evidence supporting the effect of social influence
on opinion formation processes \cite{garcia}; in consequence, to
model the agents' interactions, we follow the social impact theory
\cite{latane,lewenstein}. To give a topological structure to the
system we consider a society with $M$ agents $1\leq a\leq M$ linked
by a set of social strengths $\mathscr{S}\equiv\{\eta_{a,c}|0\leq\eta_{a,c}\in\mathbb{R}\}$,
where $\eta_{a,c}$ represents the influence agent $c$ has on the
opinion of agent $a.$ We define the neighborhood of $a$ by $\mathbb{N}_{a}=\{c|c\neq a\,\mathrm{and}\,\eta_{a,c}>0\}$
which is the set of agents $connected$ to $a$. The opinion formation
process itself is modeled by an on-line learning scenario \cite{engel},
where a set of social issues $\mathscr{L}_{P}\equiv\left\{ (\boldsymbol{\xi}_{\mu},\sigma_{B}(\boldsymbol{\xi}_{\mu})),\,\mu=1,\dots,P\right\} $
is used to define the energy of the society:
\begin{align}
E(\{\mathbf{J}_{a}\};\mathscr{L}_{P},\mathscr{S})\equiv\sum_{\mu=1}^{P}\sum_{a=1}^{M}\Theta(-\sigma_{a}(\boldsymbol{\xi}_{\mu})\sigma_{B}(\boldsymbol{\xi}_{\mu}))\nonumber \\
\left[1-\sum_{c\in\mathbb{N}_{a}}\eta_{a,c}\Theta(-\sigma_{c}(\boldsymbol{\xi}_{\mu})\sigma_{B}(\boldsymbol{\xi}_{\mu}))\right]\label{eq:full-energy}
\end{align}
where $\Theta(x)=1$ if $x>0$ and 0 otherwise. Observe that for independent
agents ($\forall a,c$ $\eta_{a,c}=0$) the energy (\ref{eq:full-energy})
is minimized with a consensus in favor of $B.$ If the social strengths
$\{\eta_{a,c}\}$ are sufficiently large, the energy is minimized
with a consensus against $B$.

Observe that the model described by (\ref{eq:full-energy}) possesses
two sources of disorder, one introduced through the set of issues
$\mathscr{L}_{P}$, and the second through the topology imposed by
$\mathscr{S}.$ In this letter we present a study on the emergence
of consensus in homogeneous, fully connected graphs (i.e. for all
index $a$, $\mathbb{N}_{a}=\{1,2,\dots,a-1,a+1,\dots,M\}$ and $\eta_{a,c}=\eta_{0}$
for all pairs $(a,c)$).

We apply the replica trick \cite{parisi} in order to compute the
expectation of the logarithm of the partition function $\overline{\log Z}=\lim_{n\to0}n^{-1}\left(\overline{Z^{n}}-1\right)$.
The average of the replicated partition function is 
\begin{eqnarray}
\overline{Z^{n}}(\beta,\eta_{0})\equiv\mathbb{E}\left[\exp\left(-\beta\sum_{\gamma=1}^{n}E(\{\mathbf{J}_{a}^{\gamma}\};\{\boldsymbol{\xi}_{\mu}\},\eta_{0})\right)\right]\label{eq:particion}
\end{eqnarray}
where the expectation $\mathbb{E}[\cdot]$ is taken over the issues
$\boldsymbol{\xi},$ the social rule $B$ and the agents' synaptic
vectors ${\bf J}_{a},$ with probabilities $\mathcal{P}(\boldsymbol{\xi})\equiv2^{-N}\prod_{k=1}^{N}(\delta_{\xi_{k},1}+\delta_{\xi_{k},-1}),$
$\mathrm{d}\mathbf{B}\mathcal{P}(\mathbf{B})=\prod_{k}\mathrm{d}B_{k}\,\delta(B_{k}-1)$
and $\mathrm{d}\mathbf{J}\mathcal{P}(\mathbf{J})\equiv\prod_{k=1}^{N}\mathrm{d}J_{k}\,\delta\left(\sum_{k=1}^{N}J_{k}^{2}-N\right)/\sqrt{2\pi\mathrm{e}}$
respectively.

By defining the order one parameters $R_{a}^{\gamma}\equiv\mathbf{J}_{a}^{\gamma}\cdot\mathbf{B}/N$,
$q_{a}^{\gamma,\rho}\equiv\mathbf{J}_{a}^{\gamma}\cdot\mathbf{J}_{a}^{\rho}/N,$
$W_{a,b}^{\gamma}\equiv\mathbf{J}_{a}^{\gamma}\cdot\mathbf{J}_{b}^{\gamma}/N,$
and $t_{a,b}^{\gamma,\rho}\equiv\mathbf{J}_{a}^{\gamma}\cdot\mathbf{J}_{b}^{\rho}/N$
and imposing the replica symmetric Ansatz, i.e. $R_{a}^{\gamma}\equiv R$,
$q_{a}^{\gamma,\rho}\equiv q,$ $W_{a,b}^{\gamma}\equiv W,$ and $t_{a,b}^{\gamma,\rho}\equiv t$
with the assumption that the overlaps $W$ and $t$ satisfy the scaling
$\tau\equiv M(W-t)\sim O(1)$ (see reference \cite{neirotti2}, equation
(3)), it is possible to demonstrate that the logarithm of $\overline{Z^{n}}$
can be decomposed in two terms, an entropic contribution:
\begin{align}
\mathcal{G}(\boldsymbol{P}) & \equiv\frac{1}{2}\left(\ln(1-q)+\frac{q-W}{1-q}+\frac{W-R^{2}}{1-q+\tau}\right)\label{eq:G}
\end{align}
with $\boldsymbol{P}=(R,q,W,\tau)$ and an energetic contribution:
\begin{align}
\mathcal{F}_{n,M}(\boldsymbol{P};\beta,\eta_{0})\equiv\frac{1}{nM}\log\left[2\int_{0}^{\infty}\!\!\!\!\!\mathcal{D}u\!\!\int\!\!\mathcal{D}w\!\!\int\!\!\prod_{a}\mathcal{D}w_{a}\right.\nonumber \\
\left.\left(\int\mathcal{D}x\mathcal{D}s\prod_{a}\left(B(x,\beta)\mathcal{H}(y_{a})+\mathcal{H}(-y_{a})\right)\right)^{n}\right]\label{eq:F}
\end{align}
where $\mathcal{D}x\equiv\mathrm{d}x\,\mathrm{e}^{-x^{2}/2}/\sqrt{2\pi}$
is the Gaussian measure, $\mathcal{H}(u)=\int_{u}^{\infty}\mathcal{D}x$
the Gardner error function, $B(x,\beta)\equiv\exp\left(\sqrt{2\beta\eta_{0}}x-\beta\right)$
and
\[
y_{a}\equiv Ru+\sqrt{W-\frac{\tau}{M}-R^{2}}w+\sqrt{q-W+\frac{\tau}{M}}w_{a}+\sqrt{\frac{\tau}{M}}s.
\]
The replicated partition function is, in the limit of large $N$:
\begin{eqnarray*}
\overline{Z^{n}}(\beta,\eta_{0})=\mathop{\mathrm{extr}}_{\boldsymbol{P}}\left\{ \exp\left[nNM\left(\mathcal{G}(\boldsymbol{P})+\alpha\mathcal{F}_{n,M}(\boldsymbol{P};\beta,\eta_{0})\right)\right]\right\} ,
\end{eqnarray*}
where $\alpha\equiv P/N$ is a measure of the volume of information
presented to the agents. Let us define the arithmetic average $\overline{y}\equiv M^{-1}\sum_{a}y_{a}$
and by using the approximation $\binom{M}{l}^{-1}\sum_{\mathbb{I}_{l}}\prod_{a=1}^{l}\mathcal{H}(-y_{i_{a}})\approx\mathcal{H}^{l}(-\overline{y})$
we can approach the replicated product in equation (\ref{eq:F}) by
\begin{align}
\prod_{a}\left(B\,\mathcal{H}(y_{a})+\mathcal{H}(-y_{a})\right) & \approx\left[B\,\mathcal{H}(\overline{y})+\mathcal{H}(-\overline{y})\right]^{M}.\label{eq:intermediate}
\end{align}
To ensure the extensivity of the energy (\ref{eq:full-energy}) we
impose the scaling $M\eta_{0}=H_{0}\sim O(1)$. Thus, by applying
a Gaussian approximation to the RHS of (\ref{eq:intermediate}) we
have that, in leading order in $M$, the energetic contribution can
be expressed as:
\begin{align}
\mathcal{F}_{n,M}(\boldsymbol{P};\beta,\eta_{0})\approx-2\sqrt{\frac{1-q}{W}}\int\frac{\mathrm{d}z}{\sqrt{2\pi}}\,\exp\left(-\frac{1-q}{W}\frac{z^{2}}{2}\right)\nonumber \\
\quad\mathcal{H}\left(-\sqrt{\frac{1-q}{W(W-R^{2})}}Rz\right)\Phi(z;\tau,q;\beta,H_{0})\label{eq:inter2}
\end{align}
where $\Phi(z;\tau,q;\beta,H_{0})$ is the minimum over $u\in(0,1)$
and $\sigma\in\mathbb{R}$ of the function 
\begin{align}
\Omega(u,\sigma,z;\tau,q;\beta,H_{0}) & \equiv\frac{1-q}{\tau}\frac{(\sigma-z)^{2}}{2}+\frac{[u-\mathcal{H}(\sigma)]^{2}}{2\mathcal{H}(\sigma)\mathcal{H}(-\sigma)}-\nonumber \\
 & -u^{2}\beta H_{0}+u\beta+O\left(\frac{\log M}{M}\right).\label{eq:oo}
\end{align}

\begin{figure}
\begin{centering}
\includegraphics[scale=0.4]{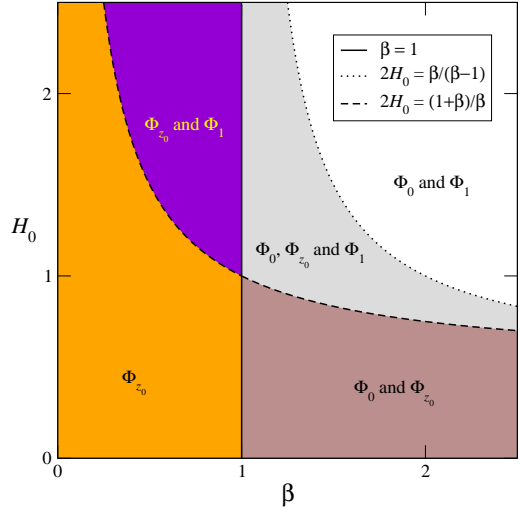}
\par\end{centering}
\caption{Distribution of the components (\ref{eq:z0}) in the plain $(\beta,H_{0})$
(color on-line).\label{fig:componentes}}
\end{figure}

For small values of $\tau$ we have that $\Phi(z;\tau,q;\beta,H_{0})=\Phi(z;\beta,H_{0})+\frac{\tau}{1-q}\hat{\Phi}(z;\beta,H_{0})+O(\tau^{2})$
for suitable functions $\Phi(z;\beta,H_{0})$ and $\hat{\Phi}(z;\beta,H_{0}).$
By defining the quantities: 
\begin{align}
a_{1} & \equiv\Theta(2H_{0}-1)\max\left\{ 0,\frac{\beta(2H_{0}-1)-1}{\beta(2H_{0}-1)}\right\} \label{eq:a1}\\
a_{2} & \equiv\min\left\{ 1,\frac{1}{\beta}\right\} \label{eq:a2}\\
a_{3} & \equiv\frac{1}{2}-\frac{\sqrt{\beta^{2}(1-H_{0})^{2}+1}-1}{2\beta(1-H_{0})}\label{eq:a3}\\
b_{0} & \equiv\Theta(a_{2}-a_{1})a_{2}+\Theta(a_{1}-a_{2})a_{3}\\
b_{1} & \equiv\Theta(a_{2}-a_{1})a_{1}+\Theta(a_{1}-a_{2})a_{3}
\end{align}
we can split the real line in three non-intersecting segments $\mathbb{D}_{z_{0}},$
$\mathbb{D}_{0}$ and $\mathbb{D}_{1}$, such that $\mathbb{D}_{z_{0}}\equiv\{x\in\mathbb{R}|b_{1}<\mathcal{H}(-x)<b_{0}\},$
$\mathbb{D}_{0}\equiv\{x\in\mathbb{R}|b_{0}<\mathcal{H}(-x)\}$ and
$\mathbb{D}_{1}\equiv\{x\in\mathbb{R}|b_{1}>\mathcal{H}(-x)\}$. In
the zeroth order of $\tau$ we have that: 
\begin{equation}
\Phi(z;\beta,H_{0})\equiv\begin{cases}
\Phi_{z_{0}}\equiv\frac{\beta\mathcal{H}(z)[1-H_{0}\mathcal{H}(z)]}{1-2\beta H_{0}\mathcal{H}(z)\mathcal{H}(-z)}- & z\in\mathbb{D}_{z_{0}}\\
\qquad-\frac{\beta^{2}\mathcal{H}(z)\mathcal{H}(-z)}{2[1-2\beta H_{0}\mathcal{H}(z)\mathcal{H}(-z)]}\\
\Phi_{0}\equiv\frac{\mathcal{H}(z)}{2\mathcal{H}(-z)} & z\in\mathbb{D}_{0}\\
\Phi_{1}\equiv\frac{\mathcal{H}(-z)}{2\mathcal{H}(z)}+\beta(1-H_{0}) & z\in\mathbb{D}_{1}.
\end{cases}\label{eq:z0}
\end{equation}
$\Phi(z;\beta,H_{0})$ is continuous in $z$ but not differentiable
at the boundaries $z=-\mathcal{H}^{-1}(a_{1})$ (between $\Phi_{1}$
to the left and $\Phi_{z_{0}}$ to the right) and $z=-\mathcal{H}^{-1}(a_{2})$
(between $\Phi_{z_{0}}$ to the left and $\Phi_{0}$ to the right)
if $a_{1}<a_{2}$ or $z=-\mathcal{H}^{-1}(a_{3})$ (between $\Phi_{1}$
to the left and $\Phi_{0}$ to the right) otherwise. In the plain
defined by the independent parameters $\beta$ and $H_{0}$ the components
$\Phi_{z_{0}},$ $\Phi_{0}$ and $\Phi_{1}$ cover the areas illustrated
in figure \ref{fig:componentes}. Observe that the component $\Phi_{z_{0}}$
appears in the sector $\mathscr{S}_{z_{0}}\equiv\{(\beta,H_{0})|\beta\leq1\,\mathrm{and}\,H_{0}\geq0\}\cup\{(\beta,H_{0})|\beta>1\,\mathrm{and}\,2H_{0}<\beta/(\beta-1)\},$
the component $\Phi_{1}$ appears in the sector $\mathscr{S}_{1}\equiv\{(\beta,H_{0})|\beta\geq0\,\mathrm{and}\,2H_{0}>(1+\beta)/\beta\}$
and the component $\Phi_{0}$ appears in the sector $\mathscr{S}_{0}\equiv\{(\beta,H_{0})|\beta\geq1\,\mathrm{and}\,H_{0}\geq0\}.$
 The fragmentation of the function $\Phi(z;\beta,H_{0})$ over the
plane $(\beta,H_{0})$ is the consequence of the interaction between
a very large number of agents and the feature responsible for the
complex behavior described in the following.

\begin{figure}
\begin{center}\includegraphics[scale=0.4]{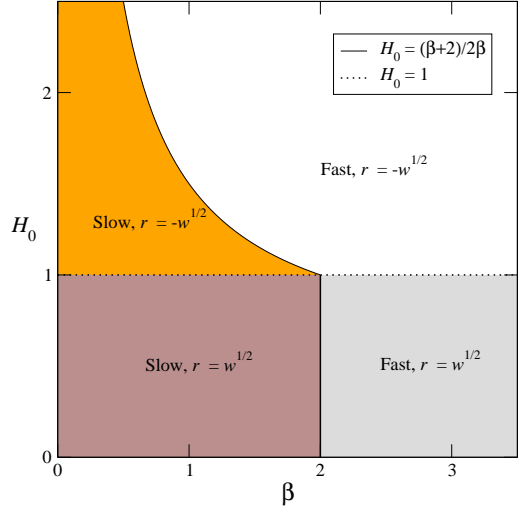}\end{center}\caption{Phase diagram of the system. For $\beta>2$ and $H_{0}<(\beta+1)/2\beta$
the volume of information the system needs to reach $r^{2}=w$ is
relatively larger. (color on-line) \label{fig:Phase-diagram}}
\end{figure}

$R$, the overlap between a typical ${\bf J}$ and ${\bf B}$, represents
the level of agreement with the social rule \emph{B.} $q$, the overlap
between synaptic vectors belonging to the same replicated agent, represents
the level of variability that remains in the space of compatible synaptic
vectors (known as the version space). $W$ is the projection of an
agent's synaptic vector in the direction of another agent's synaptic
vector within the same replicated system. If two agents $a$ and $c$
have the same overlap with $B,$ $R_{a}=R_{c}=R$ (as in the case
considered) the relationship between $R$ and $W_{a,c}$ is $W_{a,c}=R^{2}+(1-R^{2})\cos\phi$
where $\phi$ is the angle between $\mathbf{J}_{a,\perp}\equiv\mathbf{J}_{a}-R\mathbf{B}/\sqrt{N}$
and $\mathbf{J}_{c,\perp}\equiv\mathbf{J}_{c}-R\mathbf{B}/\sqrt{N}$,
which are the components of $\mathbf{J}_{a}$ and $\mathbf{J}_{c}$
perpendicular to \textbf{B} respectively. An interesting effect is
observed when we consider opinions on \emph{socially neutral issues,}
which are issues $\mathbf{S}_{0}\in\{\pm1\}^{N}$ such that ${\bf S}_{0}\cdot{\bf B}=0.$
If $\phi=\frac{\pi}{2}$ then $W_{a,c}=R^{2}$, which implies that
the opinion of an agent $a$ on $\mathbf{S}_{0}$ is independent on
the opinion agent $c\neq a$ on $\mathbf{S}_{0}$. Disregarding $\tau$
 and by defining $w\equiv(1-q)^{-1}W$ and $r\equiv(1-q)^{-1/2}R$,
we have that the free energy of the system is: 
\begin{eqnarray*}
\beta f(\alpha,\beta,H_{0}) & = & \mathop{\mathrm{extr}}_{\{r,q,w\}}\phi(r,q,w;\beta,H_{0}),
\end{eqnarray*}
where
\begin{align}
\phi(r,q,w;\beta,H_{0})\equiv-\frac{1}{2}\left(\ln(1-q)+\frac{q}{1-q}\right)+\frac{r^{2}}{2}+\nonumber \\
+2\alpha\int\frac{\mathrm{d}z}{\sqrt{2\pi w}}\exp\left(-\frac{z^{2}}{2w}\right)\mathcal{H}(-\kappa z)\Phi(z),\label{eq:free parcial}
\end{align}
where $\kappa\equiv r/\sqrt{w(w-r^{2})}.$ The conditions $\partial_{r}\phi=\partial_{q}\phi=\partial_{w}\phi=0$
imply that $q=0$ and
\begin{align}
r & =-\sqrt{\frac{2}{\pi}}\alpha\int\mathrm{d}z\,\mathcal{N}(z|0,w-r^{2})\,\Phi'(z)\label{eq:r}\\
r^{2} & =-2\alpha\int\mathrm{d}z\,\mathcal{N}(z|0,w)\,z\,\mathcal{H}(-\kappa z)\Phi'(z),\label{eq:w}
\end{align}
where $\mathcal{N}(x|\mu,\sigma^{2})\equiv\exp\left(-(x-\mu)^{2}/2\sigma^{2}\right)/\sqrt{2\pi}$
is a Gaussian distribution in $x,$ centered at $\mu,$ with standard
deviation $\sigma.$
\begin{figure}
\begin{center}\includegraphics[scale=0.38]{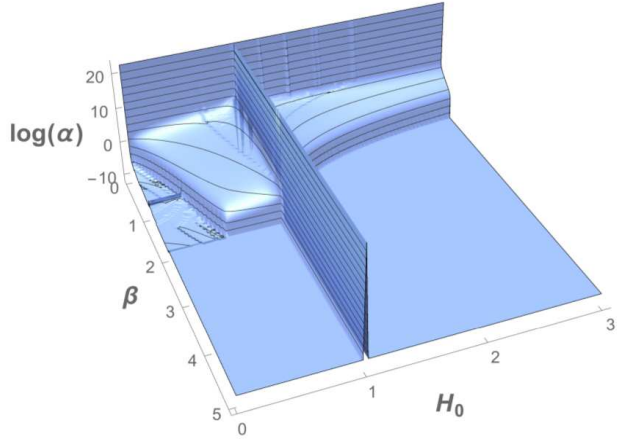}\end{center}

\caption{Logarithm of the critical volume of information $\log(\alpha_{c})$
as a function of $\beta$ and $H_{0}.$ (Color on-line)\label{fig:Loga}}
\end{figure}
Clearly, $r^{2}\leq w.$ If there is an increment of $r^{2}$ towards
$w$, the Gaussian distribution in (\ref{eq:r}) becomes sharply concentrated
at 0. Moreover, if the parameter $r^{2}$ reaches $w$ for a finite
volume of information $\alpha_{c}$ we have that
\begin{align}
\alpha_{c} & =-\sqrt{\frac{\pi}{2}}\frac{r}{\Phi'(0)}\label{eq:alfac}\\
r & =\frac{\sqrt{2\pi}}{\Phi'(0)}\int\mathrm{d}z\,\mathcal{N}(z|0,r^{2})\Theta(rz)\,z\,\Phi'(z),\label{eq:rr}
\end{align}
where
\begin{equation}
\Phi'(0)=\sqrt{\frac{2}{\pi}}\mathrm{sgn}(H_{0}-1)\begin{cases}
\frac{\beta|1-H_{0}|}{2-\beta H_{0}} & \beta<2\,\mathrm{and}\,H_{0}<\frac{\beta+2}{2\beta}\\
1 & \mathrm{otherwise}.
\end{cases}\label{eq:deriv}
\end{equation}
By defining
\begin{align*}
\mathcal{I} & \equiv-\frac{\int\mathrm{d}z\mathcal{N}(z|0,r^{2})\,\Theta((1-H_{0})z)\,\Phi''(z)}{2\int\mathrm{d}z\mathcal{N}(z|0,r^{2})\,\Theta((1-H_{0})z)\,z\,\Phi'(z)}\\
\mathcal{J} & \equiv-\sqrt{\frac{\pi}{2}}\frac{r}{\Phi'(0)}\int\mathrm{d}z\mathcal{N}(z|0,r^{2})\,\Theta((1-H_{0})z)\Phi^{(4)}(z)
\end{align*}
we have that the determinant of the Hessian when $w=r^{2}$, and $\alpha_{c}$
and $r$ are given by (\ref{eq:alfac}) and (\ref{eq:rr}) respectively
is:
\begin{equation}
|\boldsymbol{H}|=\frac{1}{4}\left[\frac{1}{r^{2}}-\frac{3}{2}\frac{\Phi'''(0)}{\Phi'(0)}+\left(1-\frac{r^{2}\Phi'''(0)}{\Phi'(0)}\right)\left(3\mathcal{I}+\mathcal{J}\right)\right]\label{eq:det}
\end{equation}
where
\[
-\frac{\Phi'''(0)}{\Phi'(0)}=\begin{cases}
\frac{1}{\pi}\frac{(12-\pi)\beta H_{0}+2\pi}{2-\beta H_{0}} & \beta<2\,\mathrm{and}\,H_{0}<\frac{\beta+2}{2\beta}\\
-\frac{12-\pi}{\pi} & \mathrm{otherwise}.
\end{cases}
\]
The determinant (\ref{eq:det}) is found to be positive for all $\beta>0$
and $H_{0}\neq1.$ Thus the solution $w=r^{2}$ at $\alpha_{c}$ given
by (\ref{eq:alfac}) with $r$ given by (\ref{eq:rr}) is stable.
A plot of the $\log(\alpha_{c})$ as a function of $\beta$ and $H_{0}$
is presented in figure \ref{fig:Loga}. From figure \ref{fig:Loga}
we observe that there is a sector of the $(\beta,H_{0})$ plane for
which the system takes a relatively long time to reach the solution
$r^{2}=w.$ This is the sector for which $z=0\in\mathbb{D}_{z_{0}}$.
In this manner we can construct the phase diagram shown in figure
\ref{fig:Phase-diagram}. 
Note that for all values of $\alpha$ larger than $\alpha_{c}$, the
equation (\ref{eq:r}) is no longer satisfied given that the minimum
occurs at the border of definition of the parameter $r$ (either $w$
or $-w$) but the derivative $\partial_{r}\phi$ is not zero. 

\begin{figure}
\begin{center}\includegraphics[scale=0.4]{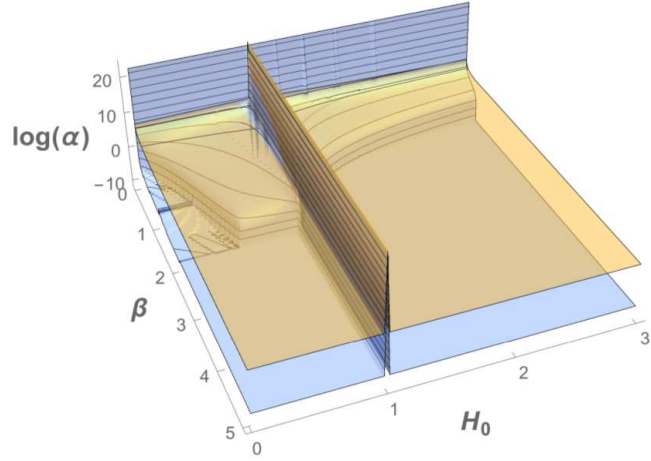}\end{center}

\caption{Comparison of the Logarithms of the critical volumes $\alpha_{c}$
and $\alpha_{d}$ functions of $\beta$ and $H_{0}.$ (Color on-line)
\label{fig:tiempos}}
\end{figure}

Most of the opinion formation process occurs for $\alpha>\alpha_{c}$.
The effective energy for $\alpha>\alpha_{c}$ can be defines as
\begin{align}
\phi_{\mathrm{eff}}(q,w;\alpha,\beta,H_{0})\equiv-\frac{\log(1-q)}{2}-\frac{q}{2(1-q)}-\frac{w}{2}+\quad\nonumber \\
+2\alpha\int\frac{\mathrm{d}z}{\sqrt{2\pi w}}\exp\left(-\frac{z^{2}}{2w}\right)\Theta\left((1-H_{0})z\right)\Phi(z).\label{eq:fparc2}
\end{align}
The new saddle point equations are
\begin{align*}
0 & =q\\
w & =2\alpha\int\mathrm{d}z\,\mathcal{N}\left(z\left|0,w\right.\right)\left(1-\frac{z^{2}}{w}\right)\Theta\left((1-H_{0})z\right)\Phi(z).
\end{align*}
There is a maximum value of $\alpha=\alpha_{d}$ such that $w=1$:
\begin{equation}
\alpha_{d}^{-1}=2\int\mathcal{D}z(1-z^{2})\Theta\left((1-H_{0})z\right)\Phi(z).\label{eq:last}
\end{equation}
The determinant of the Hessian at $\alpha_{d}$ with $q=0$ and $w=1$
is:
\[
|\boldsymbol{H}|=\frac{3}{8}-\frac{1}{8}\frac{\int\mathcal{D}z\left(3z^{2}-z^{4}\right)\Theta\left((1-H_{0})z\right)\Phi(z)}{\int\mathcal{D}z(1-z^{2})\Theta\left((1-H_{0})z\right)\Phi(z)}
\]
which is positive for all values of $\beta$ and $H_{0.}$ As it is
shown in figure \ref{fig:tiempos}, $\alpha_{c}<\alpha_{d}$ for all
values of $\beta>0$ and $H_{0}\neq1.$ 

\emph{Discussion:} We presented a model for the opinion formation
process in a society of interacting agents, represented by binary
perceptrons, in the presence of a social field $B$. The field is
the result of many opinion formation processes prior to the current
one and provides the socially acceptable position on which issues'
opinions are formed. 

Although we worked in a fully connected graph with non directed links,
we observed the asymptotic formation of a consensus for all temperatures
and values of the interaction, with the exception of the lines $\beta=0$
and $H_{0}=1$. On the line $\beta=0$ consensus is not achieved due
to large energy fluctuations in the system. At $H_{0}=1$ competing
attitudes towards following either $B$ or neighboring agents cancel
each other and consensus is never reached.

The solution to the saddle point equations for the energy (\ref{eq:free parcial})
reveal the following behavior: Firstly, the overlap $q=0$ whatever
the value of $\alpha.$ This indicates that a maximum of variability
is kept in the version space. The first milestone in the opinion formation
process is reached at $\alpha_{c}(\beta,H_{0}),$ which is the volume
of information at which $R^{2}=W$. From this point onwards the agents
approach consensus disregarding the opinion of their peers on socially
neutral issues. The majority of the opinion formation process occurs
for volumes $\alpha_{c}<\alpha<\alpha_{d},$ where the effective energy
of the system is described by (\ref{eq:fparc2}). At $\alpha_{d}$
$W=1$ and consensus is reached. For values of $H_{0}<1$ $R=\sqrt{W}=1$
and all the agents follow the\emph{ status quo} imposed by \emph{B}.
For $H_{0}>1$ $R=-\sqrt{W}=-1$ and the consensus is against \emph{B.
}In both cases the agents reach a consensus following a path that
maximizes the diversity of opinions in the only manner allowed: by
developing independent attitudes towards socially neutral issues.
$\alpha$ is a time-like parameter, thus the reported $\alpha_{c}$
and $\alpha_{d}$ can be considered as characteristic times of the
model, which, for a fully connected system, are expected to be shorter
than the characteristic times of a system defined on a more realistic
graph \cite{li,baxter}. 

As it is expected from a mean field approximation \cite{krapivsky,soulier},
phenomena associated to the correlation length of the system (like
the presence of clusters reported in \cite{neirotti1,shao}), cannot
be addressed within this framework. To do so we will need to consider
more realistic graph topologies, particularly by introducing non-symmetric
interaction (directed graphs) \cite{sanchez} and connectivity dynamics
\cite{gross,nardini} which facilitates the exchange of information
between agents \cite{toscani,fortunato}. 
\begin{acknowledgments}
The author would like to acknowledge discussions with Dr R. C. Alamino
in the early stages of this work. The advice of L. E. Neirotti is
also warmly appreciated.
\end{acknowledgments}

\end{document}